\documentclass[11pt]{article}

\usepackage{amsfonts}
\usepackage{amssymb}
\usepackage{amscd}
\usepackage{amsmath}

\newcommand{\BEQ}{\begin{equation}}
\newcommand{\EEQ}{\end{equation}}
\def\bea{\begin{eqnarray}}
\def\eea{\end{eqnarray}}
\def\nn{\nonumber}

\newtheorem{Th}{Theorem}
\newtheorem{lem}{Lemma}

\newtheorem{Rem}{Remark}

\def\bea{\begin{eqnarray}}
\def\eea{\end{eqnarray}}

\def\bes{\begin{equation*} \begin{split}}
\def\ees{\end{split} \end{equation*}}

\def\O{{\cal O}}
\def\Z{\mathbb{Z}}
\def\C{{\mathbb{ C}}}

\def\A{\mathcal{A}}

\def\g{\mathfrak{gl}_n}

\def\l{\lambda}
\def\m{\mu}

\def\a{\alpha}
\def\p{\partial_z}
\def\ga{\gamma}

\begin{document}
\begin{titlepage}
\hfill ITEP-TH-03/08

\vskip 3.0cm
{\LARGE \bf {Bethe ansatz and }}
\vskip 3mm
{\LARGE \bf {Isomonodromic deformations}}

\vskip 1.0cm \centerline{D. Talalaev \footnote{E-mail:
talalaev@itep.ru} }
\vskip 1.0cm \centerline{\sf Institute for Theoretical and Experimental
Physics \footnote{ITEP, 25 B. Cheremushkinskaya, Moscow, 117259, Russia.}}
\vskip 1.0cm 

\begin{abstract}
We study symmetries of the Bethe equations for the Gaudin model appeared naturally in the framework of the geometric Langlands correspondence under the name of Hecke operators and under the name of Schlesinger transformations in the theory of isomonodromic deformations, and particularly in the theory of Painlev\'e transcendents.
\end{abstract}


\end{titlepage}

\section*{Introduction}
The Gaudin model is one of the first nontrivial examples of quantum integrable systems which shows deep connection of problems of mathematical physics and such modern fundamental mathematical questions as the Langlands correspondence. For $\C P^1$ with marked points the geometric Langlands correspondence is an interpretation of the separation of variables in quantum Gaudin model.
The basic purpose of this work is to specify the Langlands correspondence concerning the problem of solution of a quantum model. We use the form of the Bethe equations which expresses a condition 
for the spectrum of the Gaudin model in terms of the monodromy properties of certain differential operator with regular singularities (called the $G$-oper). Further we construct the discrete monodromy preserving transformations known as Schlesinger or Hecke transformations. It appears, that they determine also an action on rational solutions of the G-oper, and therefore on eigenvectors of the Gaudin model.
It is necessary to mention that such transformations act by changing the highest weights in representations and also the inhomogeneity parameters of the model.

The exposition is organized as follows: in section \ref{definition} the Gaudin model is defined, the traditional version of the Bethe ansatz and its formulation in terms of monodromy properties of specific differential equations is described. In section \ref{matrix} the "matrix" form of the monodromy condition equivalent to the Bethe equations is constructed. In section \ref{schl} the Schlesinger transformations are investigated. The last section is devoted to some relations of this subject with other problems.

{\bf Acknowledgments.~} The part of this work has been done during the stay of the author at LPTHE, the author is grateful for the stimulating atmosphere. The proof of the theorem \ref{back} was obtained during the stay of the auther at LAREMA.
This work was partially supported by the Federal Nuclear Energy Agency of Russia, the RFBR grant 07-02-00645, the grant of Support for the Scientific Schools 3035.2008.2, and the fund "Dynasty". Author would like to thank O. Babelon, B. Feigin, A. Gorodentsev, I. Krichever and V. Rubtsov for useful remarks.

\section{Bethe ansatz for the Gaudin model}
\label{definition}
\subsection{Gaudin model}
The Lax operator of this model is a rational function with simple poles:
\bea
\label{main-part-Lax-Gaud}
L(z)=\sum_{i=1}^N
\frac{\Phi_{i}}{z-z_i}\nn,
\eea
with the residues $\Phi_i\in Mat_n\otimes \bigoplus\g \subset Mat_n\otimes
U(\g)^{\otimes N}$  defined by the formula:
\bea \label{Phi-Gaudin}
\Phi_i=\sum_{kl}E_{kl}\otimes e_{kl}^{(i)}
\eea
where $E_{kl}$ form the  standard basis
in $Mat_n$ and $e_{kl}^{(i)}$ form a basis in the $i$-th copy of the Lie algebra $\g.$
It is the same as to say that the quantum algebra is 
$$
 \A = U(\g)^{\otimes N}
$$
and the commutation relations for the Lax operator are
$$
[L(z)\otimes 1,1\otimes L(u)]=[R_{12}(z-u),L(z)\otimes 1+1\otimes L(u)]\in
Mat_n^{\otimes 2}\otimes \A
$$
Let us now define the ``quantum'' determinant of a matrix with non commutative entries as the completely symmetrized determinant
\bea
"det"(B)=\frac 1 {n!}
\sum_{\tau,\sigma\in
\Sigma_n}(-1)^{\tau\sigma}B_{\tau(1),\sigma(1)}\ldots
B_{\tau(n),\sigma(n)}\nn
\eea
Then the quantum spectral curve is defined as follows
\bea \label{qchar}
"det"(L(z)-\partial_z)
=\sum_{k=0}^n QI_k(z)\p^{n-k}
\eea
{\bf Theorem \cite{T04}} {\it The coefficients $QI_k(z)$ commute
\bea
[QI_k(z),QI_m(u)]=0\nn
\eea
and quantize the classical Gaudin hamiltonians.}

{\Rem By the same formula in \cite{CT06-1} it was constructed a commutative subalgebra in
$U(\g)[t]/{t^N},$ $U(\g)[t]$ and the center of the universal enveloping affine
algebra on the critical level .}

\subsection{Bethe ansatz}
\subsubsection{$\mathfrak{sl}_2$ case}
Let us consider the Gaudin model for the $\mathfrak{sl}_2$ case. The Lax operator can be expressed by the formula
$$
L=\left(\begin{array}{cc}
A(z) & B(z)\\
C(z) & D(z)
\end{array}\right)
=\sum_{i=1}^N \frac {\Phi_i}{z-z_i}
$$ where 
$$ \Phi_i=\left(
\begin{array}{cc}
h_i/2 & e_i \\
f_i &-h_i/2
\end{array} \right)
$$
Let us also fix a finite dimensional representation
$V_\l=\otimes V_{\l_i}$ which is the tensor product of highest weight representations for each factor $U(\mathfrak{sl}_2)$ of the quantum algebra. The vacuum vector is
 $|vac>=\otimes_i |0>_{i}$ where the vectors $|0>_i$ are defined by $e_i|0>_i=0$ and $h_i|0>=\l_i |0>.$
The quantum characteristic polynomial in this case is
$$"det"(L(z)-\p)=-\frac 1 2 Tr L^2(z) +\p^2$$
$$
Tr L^2(z)=\sum_i \frac {Tr \Phi_i^2}{(z-z_i)^2}+2\sum_{i<j}\frac {Tr(\Phi_i \Phi_j)}{(z-z_i)(z-z_j)}
$$
The coefficients at the double poles are central
$$Tr \Phi_i^2=h_i^2/2+e_i f_i +f_i e_i=c_i^{(2)}$$
and take values $1/2 \l_i(\l_i+2)$ in representation $V_{\l}.$
Hence
$$"det"(L(z)-\p)=\p^2-\frac 1 2\sum_i \frac {c_i^{(2)}}{(z-z_i)^2}- \sum_i \frac {H_i} {z-z_i}$$
where the residues $H_i$ are the Gaudin hamiltonians 
$$H_i=\sum_{i\ne j}\frac {h_i h_j/2+e_i f_j+e_j f_i}{z_i-z_j}$$

\subsubsection{Bethe vector}
The Bethe ansatz method supposes that a common eigenvector $\Omega$  in $V_{\l}$
\bea
H_i \Omega = \chi_{\Omega}(H_i) \Omega
\eea
has the form 
$$\Omega=\prod_{j=1}^M C(\m_j)|vac>$$ 
for some values of parameters $\m_j.$ This condition implies the following system of algebraic equation on parameters $\m_j$ (the Bethe system)
\bea
\label{b}
-\frac 1 2 \sum_{i}\frac {\l_i}{\m_j-z_i}+\sum_{k\ne j} \frac 1 {\m_j-\m_k}=0~~~ j=1,\ldots, M
\eea
The eigenvalues take the form
$$
\chi_{\Omega}(H_i) =-\l_i\left(\sum_j \frac 1 {z_i-\m_j}-\frac 1 2 \sum_{j\ne i}\frac {\l_j}{z_i-z_j}\right)
$$

\subsection{``Null-monodromic'' form}
Let
\bea
\label{SL}
\Psi(z)=\prod (z-z_i)^{-\l_i/2}\prod (z-\m_j)
\eea
be a solution of the equation
\bea
\label{baxter}
"det"(L(z)-\p)\Psi(z)=\left(\p^2-\frac 1 2\sum_i \frac {c_i^{(2)}}{(z-z_i)^2}- \sum_i \frac {H_i} {z-z_i}\right)\Psi(z)=0
\eea
supposing that $H_i$ are scalars.
This is equivalent to the following system of equations
\bea
\frac 1 {z-z_i}: && -\l_i\left(\sum_{j\ne i}\frac {-\l_i/2}{z_i-z_j}+\sum_j \frac 1 {z_i-\m_j}\right)=H_i\nn\\
\frac 1 {z-\m_j}: &&  -\frac 1 2 \sum_{i}\frac {\l_i}{\m_j-z_i}+\sum_{k\ne j} \frac 1 {\m_j-\m_k}=0\nn
\eea
which is exactly the system of Bethe equations and the conditions on eigenvalues.
Remarkably the second solution of the equation \ref{baxter} has the same behavior at $z=z_i$, that is at most half integer exponents. This imply that the monodromy of the fundamental solution lies in $\Z/2\Z.$ There is also an inverse statement
\\
{\bf Theorem \cite{MTV}} {\it 
The equation
\bea
\left(\partial^2-\frac 1 4 \sum_i \frac {\l_i(\l_i+2)}{(z-z_i)^2}- \sum_i \frac {\nu_i} {z-z_i} \right)\Psi(z)=0
\eea
with $\nu_i\in \C$
has monodromy $\pm 1$
iff $\nu_i$ are equal to $\chi_{\Omega}(H_i)$ for some common eigenvector $\Omega$ of the Gaudin hamiltonians in the representation $V_{\l}.$}

\section{Matrix form}
\label{matrix}
\subsection{Scalar  versus matrix $G$-opers}
Let us now consider a connection in the trivial rank $2$ bundle on the punctured disc of the form:
\bea
A(z)=\left(\begin{array}{cc}
a_{11}(z) & a_{12}(z)\\
a_{21}(z) & a_{22}(z)
\end{array}\right)=\sum_{i=1}^k\frac {A_i}{z-z_i}
\eea
where the residues satisfy the following conditions
\bea
\label{norm2}
Tr(A_i)=0;~~~Det(A_i)=-d_i^2;~~~\sum_{i}A_i=\left(\begin{array}{cc}
\kappa & 0\\
0 & -\kappa
\end{array}\right).
\eea
Let us now consider the Fuchsian system
\bea
\label{aux}
(\p-A(z))\Psi(z)=0
\eea
which is equivalent to the following
\bea
\psi_1'&=& a_{11}\psi_1+a_{12}\psi_2\nn\\
\psi_2'&=& a_{21}\psi_1+a_{22}\psi_2\nn
\eea
This system implies
\bea
\psi_1''=a_{12}'/a_{12}\psi_1'+u\psi_1\nn
\eea
where
\bea
u=a_{11}'+a_{11}^2-a_{11}a_{12}'/a_{12}+a_{12}a_{21}\nn
\eea
Then if one passes to $\Phi=\psi_1/\chi$ where $\chi=\sqrt{a_{12}}$
one obtains the equation
\bea
\Phi''+U\Phi=0\nn
\eea
where
\bea
U=\chi''/\chi-a_{12}'/a_{12}\chi'/\chi-u
\eea
More precisely this implies
\bea
U=\frac 1 2 \left(\frac {a_{12}'}{a_{12}}\right)'-\frac 1 4 \left(\frac {a_{12}'}{a_{12}}\right)^2+
a_{11}\frac {a_{12}'}{a_{12}}-a_{11}'-a_{11}^2
-a_{12}a_{21}
\eea
Suppose that $a_{12}(z)$ has distinct zeros
\bea
a_{12}(z)=c \frac {\prod_{j=1}^{k-2}(z-w_j)}{\prod_{i=1}^k(z-z_i)}\nn
\eea
Let us note that the number of zeros is in agreement with the normalization condition \ref{norm2}.
Hence 
\bea
\frac {a_{12}'}{a_{12}}=\sum_{j=1}^{k-2}\frac 1 {z-w_j} -\sum_{i=1}^{k}\frac 1 {z-z_i}
\eea
The expression for the potential takes the following form
\bea
U=\sum_{j=1}^{k-2}\frac {-3/4}{(z-w_j)^2}+\sum_{i=1}^{k}\frac {1/4+det A_i}{(z-z_i)^2}+
\sum_{j=1}^{k-2}\frac {H_{w_j}}{z-w_j}+\sum_{i=1}^{k}\frac {H_{z_i}}{z-z_i}
\eea
where
\bea
H_{w_j}&=&a_{11}(w_j)+\frac 1 2\left(\sum_{i\ne j}\frac 1 {w_j-w_i}-\sum_{i}\frac 1 {w_j-z_i}\right)\nn\\
H_{z_i}&=&\left(\frac 1 2 + a_{11}^i\right)\sum_{j}\frac 1 {z_i-w_j}-\sum_{j\ne i}
\frac {Tr(A_i A_j)+a_{11}^i+a_{11}^j+1/2}{z_i-z_j}\nn
\eea
Let us note that the coefficients at $(z-z_i)^{-2}$ can be rewritten in the form 
\bea 
1/4+det A_i=(1/2-d_i)(1/2+d_i)
\eea
Further we will identify this with the values of Casimir elements. In this way $\l_i=2d_i-1.$

\subsection{Dual $G$-oper}
As it was shown in previous calculation the matrix connection is related to the Sturm-Liouville operator with additional regular singularities at points $w_j.$ The similar consideration for the second component $\Psi_2$ can produce another scalar differential operator of the second order with poles at $z_i$ and additional points $\widetilde{w}_j$ defined by the formula
\bea
a_{21}(z)=\widetilde{c}\frac {\prod_{j=1}^{k-2}(z-\widetilde{w}_j)}{\prod_{i=1}^k(z-z_i)}\nn
\eea
We will call the corresponding Sturm-Liouville operator 
\bea
\p^2-\widetilde{U}\nn
\eea
the dual $G$-oper
where 
\bea
\widetilde{U}=\sum_{j=1}^{k-2}\frac {-3/4}{(z-\widetilde{w}_j)^2}+\sum_{i=1}^{k}\frac {1/4+det A_i}{(z-z_i)^2}+
\sum_{j=1}^{k-2}\frac {H_{\widetilde{w}_j}}{z-\widetilde{w}_j}+\sum_{i=1}^{k}\frac {\widetilde{H}_{z_i}}{z-z_i}
\eea

\subsection{Pull back}
In this section we construct the inverse map, namely from the scalar $G$-oper, which has trivial monodromy in the sense of the Bethe equations, we construct a rank $2$ connection of the form \ref{aux}, such that its horisontal sections have monodromy in $\Z/2\Z$.

Let us consider the solution for the matrix linear equation \ref{aux}
\bea
(\p-A(z))\Psi=0\nn
\eea
in the following ansatz 
\bea
\label{ML}
\psi_l=\prod_{i=1}^k(z-z_i)^{-s_i}\phi_l(z)
\eea
where 
\bea
\phi_1=\prod_{j=1}^{M} (z-\ga_j)\nn\\
\phi_2/\phi_1=\sum_{j=1}^M\frac {\alpha_j}{z-\ga_j} 
\label{new_param}
\eea
Let us rewrite the system \ref{aux} with respect to the new parameterization \ref{new_param}
\bea
\p \psi_1/\psi_1=a_{11}+a_{12}\phi_2/\phi_1\label{eq3}\\
\p \psi_1/\psi_1 (\phi_2/\phi_1)+ \p(\phi_2/\phi_1)=a_{21}+a_{22}\phi_2/\phi_1
\label{eq4}
\eea
Let us represent these equations in a more explicit form:
\bea
\label{eq5}
-\sum_i \frac {s_i}{z-z_i}+\sum_j\frac 1 {z-\ga_j}=\sum_i\frac {a_{11}^i}{z-z_i}+\sum_i\frac {a_{12}^i}{z-z_i}\sum_j\frac {\a_j}{z-\ga_j}\\
\label{eq6}
\left(-\sum_i \frac {s_i}{z-z_i}+\sum_j\frac 1 {z-\ga_j}\right)\sum_j\frac {\a_j}{z-\ga_j}-\sum_j\frac {\a_j}{(z-\ga_j)^2}=\sum_i\frac {a_{21}^i}{z-z_i}-\sum_i\frac {a_{11}^i}{z-z_i}\sum_j\frac {\a_j}{z-\ga_j}
\eea
Further we find the conditions, that the residues of both sides of \ref{eq3},\ref{eq4} at $z=z_i$ coincide:
\bea
\label{res1}
-s_i &=& a_{11}^i +a_{12}^i \sum_j\frac {\a_j}{z_i-\ga_j}\\
\label{res2}
-\sum_j\frac {\a_j}{z_i-\ga_j}s_i &=& a_{21}^i -a_{11}^i \sum_j\frac {\a_j}{z_i-\ga_j}
\eea
These equations together with the trace condition $a_{11}^i+a_{22}^i=0$ imply that $s_i$ are the eigenvalues of $A_i,$ in particular the choice $s_i=d_i$ is appropriate. Let us consider the behavior at poles $z=\ga_j.$ Remarkably the poles of the second order at these points in equation \ref{res2} reduce. Let us calculate teh residues of both sides of the equations \ref{res1} and \ref{res2} 
\bea
\label{eq7}
1&=&\a_j \sum_i\frac {a_{12}^i}{\ga_j-z_i}\\
\label{eq8}
\a_j\left(-\sum_i\frac {s_i}{\ga_j-z_i}+\sum_{i\ne j}\frac 1 {\ga_j-\ga_i}\right)+\sum_{i\ne j}\frac {\a_i}{\ga_j-\ga_i}&=&-\a_j\sum_{i}\frac {a_{11}^i}{\ga_j-z_i}
\eea
Let us recall the normalizing condition on the residue at $\infty$
\bea
\label{lin2}
\sum_{i=1}^k a_{12}^i&=&0\\
\label{lin2'}
\sum_{i=1}^k a_{21}^i&=&0
\eea
Let us note that the choice of poles for the Sturm-Liouville operator fixes the zeros of the rational function $a_{12}(z),$ which in turn is defined up to a constant:
\bea
a_{12}(z)=c \frac {\prod_{j=1}^{k-2}(z-w_j)}{\prod_{i=1}^k(z-z_i)}\nn
\eea
Then the condition \ref{lin2} fulfills automatically. The coefficients $a_{12}^i$ can be expressed by the formula
\bea
\label{a12}
a_{12}^i=c\frac {\prod_{j=1}^{k-2}(z_i-w_j)}{\prod_{j\ne i}^k(z_i-z_j)}
\eea
The coefficients $a_{11}^i$ can also be expressed as follows in virtue of \ref{res1}
\bea
\label{a11}
a_{11}^i=-s_i-c\frac {\prod_{j=1}^{k-2}(z_i-w_j)}{\prod_{j\ne i}^k(z_i-z_j)}\sum_l\frac {\a_l}{z_i-\ga_l}
\eea
Let us substitute the expressions for $a_{12}^i$ and $a_{11}^i$ in equations \ref{eq7}, \ref{eq8}, and express then $\a_j$ from the first one and substitute to the other:
\bea
&&\frac {\prod_i(\ga_j-z_i)}{\prod_i(\ga_j-w_i)}\left(-\sum_k\frac {2s_k}{\ga_j-z_k}+\sum_{k\ne j}\frac 1 {\ga_j-\ga_k}+\sum_{k,m}\frac {\prod_l(z_k-w_l)\prod_{s\ne k}(\ga_m-z_s)}{\prod_{l\ne k}(z_k-z_l)\prod_s(\ga_m-w_s)(\ga_j-z_k)}\right)\nn\\
&+&\sum_{k\ne j}\frac{\prod_i(\ga_k-z_i)}{\prod_i(\ga_k-w_i)(\ga_j-\ga_k)}=0\nn
\eea
Let us pass to the equivalent form dividing by $\frac {\prod_i(\ga_j-z_i)}{\prod_i(\ga_j-w_i)}$
\bea
&&-\sum_k\frac {2s_k}{\ga_j-z_k}+\sum_{k\ne j}\frac 1 {\ga_j-\ga_k}+\sum_{k,m}\frac {\prod_l(z_k-w_l)\prod_{s\ne k}(\ga_m-z_s)}{\prod_{l\ne k}(z_k-z_l)\prod_s(\ga_m-w_s)(\ga_j-z_k)}\nn\\
&+&\sum_{k\ne j}\frac{\prod_i(\ga_k-z_i)}{\prod_i(\ga_k-w_i)(\ga_j-\ga_k)}\frac {\prod_m(\ga_j-w_m)}{\prod_m(\ga_j-z_m)}=0\nn
\eea
Consider the left hand side of the equality as a rational function $F(\ga_j)$ and calculate its simple fraction decomposition at poles $z_k,~w_k,~\ga_k$ and $\infty.$ It turns out that this decomposition takes the form:
\bea
F(\ga_j)=-\sum_k\frac {2s_k-1}{\ga_j-z_k} -\sum_k\frac 1 {\ga_j-w_k}+2\sum_{i\ne j}\frac 1 {\ga_j-\ga_i}
\eea
Hence, the explored equality is equivalent to one of the Bethe equation. Let us demonstrate, for example, the calculation of the residue at $\ga_j=w_i$
\bea
Res_{\ga_j=w_i}F(\ga_j)&=&\sum_{k}\frac {\prod_l(z_k-w_l)\prod_{s\ne k}(w_i-z_s)}{\prod_{l\ne k}(z_k-z_l)\prod_{s\ne i}(w_i-w_s)(w_i-z_k)}\nn\\
&=&\frac {\prod_s(w_i-z_s)}{\prod_{s\ne i}(w_i-w_s)}\sum_k\frac {\prod_l(z_k-w_l)}{\prod_{l\ne k}(z_k-z_l)(w_i-z_k)^2}
\eea
Let us note that the expression on the right is of the form
\bea
(Res_{z=w_i}\Phi(z))^{-1}\sum_{k}Res_{z=z_k}\Phi(z)\nn
\eea
where
\bea
\Phi(z)=\frac {\prod_l(z-w_l)}{\prod_l(z-z_l)(z-w_i)^2}
\nn
\eea
and hence is equal to $-1.$

Then the sufficient condition expresses in the following 
{\Th \label{back}
If the set of numbers $\ga_i$ where $i=1,\ldots,M$ satisfy the  system of Bethe equations with parameters: the set of poles is $z_1,\ldots, z_k$ and $w_1,\ldots,w_{k-2}$ with the highest weghts $2s_1-1,\ldots,2s_k-1$ and $1,\ldots,1$ correspondingly, then the vector
\bea
\Psi=\prod_{i=1}^k(z-z_i)^{-s_i}
\left(\begin{array}{c}
\phi_1(z)\\
\phi_2(z)
\end{array}\right)
\eea
where
\bea
\phi_1=\prod_{j=1}^{M} (z-\ga_j)\nn\\
\phi_2/\phi_1=\sum_{j=1}^M\frac {\alpha_j}{z-\ga_j} 
\eea
and the coefficients $\a_j$ given by the expressions
 \bea
 \a_j=\frac {\prod_i(\ga_j-z_i)}{\prod_i(\ga_j-w_i)}
 \eea
solves the matrix linear problem \ref{aux} where the connection coefficients are defined by
 \bea
 a_{12}^i&=&\frac{\prod_j(z_i-w_j)}{\prod_{j\ne i}(z_i-z_j)} 
 \eea
 the coefficients $a_{11}^i$ and $a_{21}^i$ are defined from \ref{res1}, \ref{res2}. The normalizing conditions \ref{norm2} hold.
 }
 
 ~\\
 {\bf Proof.~ } Properly, we need to that the normalization condition \ref{lin2'} does not depend on the choice of the parameter $c$, in particular can be taken to be equal $1$. Indeed, in virtue of \ref{res1}, \ref{res2} we obtain:
\bea
a_{21}^i=-\left(2s_i\sum_j\frac {\a_j}{z_i-\ga_j}+\frac{\prod_j(z_i-w_j)}{\prod_{j\ne i}(z_i-z_j)}\left(\sum_j\frac {\a_j}{z_i-\ga_j}\right)^2\right)
\eea
We need to prove that 
\bea
\label{eq11}
\sum_i 2s_i\sum_j\frac {\a_j}{z_i-\ga_j}+\sum_i \frac{\prod_j(z_i-w_j)}{\prod_{j\ne i}(z_i-z_j)}\left(\sum_j\frac {\a_j}{z_i-\ga_j}\right)^2=0
\eea
The first summand of \ref{eq11} can be transformed to the following expression using the Bethe equations:
\bea\label{eq11'}
\sum_i 2s_i\sum_j\frac {\a_j}{z_i-\ga_j}&=&\sum_j \a_j\sum_i\frac {2s_i}{z_i-\ga_j}\nn\\
&=&\sum_j \a_j\left(-\sum_i\frac 1 {\ga_j-z_i}+\sum_i\frac 1 {\ga_j-w_i}-2\sum_{i\ne j}\frac 1 {\ga_j-\ga_i}\right)
\eea
Let us simplify the second summand of \ref{eq11} changing the summation order
\bea
\label{eq12}
\sum_{m\ne l}\a_m \a_l\sum_i\frac {\prod_j (z_i-w_j)}{\prod_{j\ne i}(z_i-z_j)(z_i-\ga_m)(z_i-\ga_l)}+\sum_{m}(\a_m)^2 \sum_i\frac {\prod_j (z_i-w_j)}{\prod_{j\ne i}(z_i-z_j)(z_i-\ga_m)^2}
\eea
Considering the second summand of \ref{eq12} let us emphasize that
\bea\label{eq14}
\sum_i\frac {\prod_j (z_i-w_j)}{\prod_{j\ne i}(z_i-z_j)(z_i-\ga_m)^2}=-\partial_{\ga_m}\Phi^1(\ga_m)
\eea
where
\bea
\Phi^1(\ga_m)=\sum_i\frac {\prod_j (z_i-w_j)}{\prod_{j\ne i}(z_i-z_j)(z_i-\ga_m)}=\frac {\prod_j(\ga_m-w_j)}{\prod_j(\ga_m-z_j)}
\eea
Hence the expression \ref{eq14} takes the form:
\bea
-\frac {\prod_j(\ga_m-w_j)}{\prod_j(\ga_m-z_j)}\left(\sum_s\frac 1 {\ga_m-w_s}-\sum_s \frac 1 {\ga_m-z_s}\right)
\eea
which reduces the corresponding part of \ref{eq11'}.
Let us consider the first summand of \ref{eq12}, it can also be simplified:
\bea
\sum_i\frac {\prod_j (z_i-w_j)}{\prod_{j\ne i}(z_i-z_j)(z_i-\ga_m)(z_i-\ga_l)}=\frac{\prod_j(\ga_l-w_j)}{\prod_j(\ga_l-z_j)(\ga_m-\ga_l)}-\frac {\prod_j(\ga_m-w_j)}{\prod_j(\ga_m-z_j)(\ga_m-\ga_l)}
\eea
Inserting this expression in \ref{eq12} we finish the proof $\square$

\section{Schlesinger transformations}
\label{schl}
There is a discrete group of transformations which preserve the form of the connection \ref{aux} and moreover do not change the class of monodromy representation. However they change the characteristic exponents at singular points by half-integers in a specific manner. Such transformations are called Schlesinger, Hecke or  B\"aklund transformations in different contexts. This type of transformation has simple geometric interpretation.

\subsection{Action on bundles}
Let us first of all consider holomorphic bundles on $\C P^1.$ Due to the Birkhof-Grotendieck theorem any holomorphic bundle on $\C P^1$ of rank $n$ is isomorphic to $\O(k_1)\oplus\ldots\oplus \O(k_n)$ for some tuple $(k_1,\ldots,k_n)$ which is called the type of the bundle and is defined up to the symmetric group action.

Let us now consider a covering of $\C P^1$ which consist of $U_{\infty}$ - a disc around $\infty$ not containing $z=z_i,~i=1,\ldots,N$ and $U_{0}=\C P^1 \backslash \{\infty\}.$ We will work with $rk=2$ holomorphic bundles and parameterize them by the gluing function $G(z)$ - holomorphic invertible  function in $U_{0}\cap U_\infty$ with values in $GL(2).$ We say that the pair $S_{\infty}(z)\in \O^{(2)}(U_{\infty})$ and  $S_{0}(z)\in \O^{(2)}(U_{0})$ defines a global section if $S_{0}(z)=G(z)S_{\infty}(z).$

All transformations will be described by some change of the gluing function. Let us consider such a transformation which multiplies the gluing function on the left by 
\bea
G_s(z)=G_s\left(\begin{array}{cc}
z-z_s & 0\\
0 & 1
\end{array} \right)G_s^{-1}
\eea
for some constant matrix $G_s$ and some point $z_s\in U_{0}.$
{\Rem The action on gluing functions can be brought to the action on isomorphism classes of bundles making the choice of constant matrix $G_s$ dependent on the trivialization, more precisely if one change the trivialization in $U_{0}$ by $T(z)$ one should change the matrix $G_s$ by $T(z_s) G_s.$ This is clear from the remark \ref{inv_def} on the invariant definition. }

The case of our interest will be the composition of two such transformations at different points.
\begin{lem}
The composition of two transformations given by $G_i(z) G^{-1}_j(z)$ applied to the trivial bundle give also the trivial one for generic choice of constant matrices $G_i,~G_j.$
\end{lem}
{\bf Proof~~} Our goal is to perform the decomposition $G(z)=G_i(z) G^{-1}_j(z)=G_{ij}(z)G_{\infty}(z)$
where $G_{ij}(z),~G_{\infty}(z)$ are invertible respectively in $U_{0},~U_{\infty}.$
The general argument here is the observation about the dimension of cohomology spaces in families at generic point. Indeed, for particular choice of $G_i^{-1}G_j=1$ one obtains the trivial bundle which is semistable and hence minimizes the dimension of $H^0(End(V))$ for $V$ of degree $0$ which is the case. Due to this the trivial bundle is generic in the family of bundles given by different $G$.

However we provide here an explicit demonstration in spirit of the decomposition lemma of \cite{bol}. 
Let us decompose the product 
\bea
G(z)=G_i
\left(\begin{array}{cc}
z & 0 \\
0 & 1
\end{array}\right) G_i^{-1} G_j 
\left(\begin{array}{cc}
(z-1)^{-1} & 0 \\
0 & 1
\end{array}\right) G_j^{-1}
\eea
where we use the following notations
\bea
G_i=\left(\begin{array}{cc}
1 & x_i\\
y_i & 1
\end{array}\right)
\eea
into the product 
\bea
G(z)=G_{ij}(z) G_\infty(z)\nn
\eea
where 
$G_{ij}(z),~G_{\infty}(z)$ are holomorphic invertible functions in $U_{0},~U_{\infty}$ respectively.
Traditional calculations provide a decomposition of the form
\bea
G_\infty(z)=
\left(\begin{array}{cc}
\frac {z(1-x_j y_i)(1-x_j y_j)-x_j(y_i-2y_j-x_j y_i y_j)}{(1-2 x_j y_i+x_j y_j)(1-x_j y_i)
(1-x_j y_j)(z-1)}& -\frac {x_j}{(1-x_j y_i)(1-x_j y_j)(z-1)} \nn\\
\frac {y_i-2y_j+x_j y_i y_j}{(1-x_j y_i)(1-2 x_j y_i+ x_j y_j)}& \frac {1}{1-x_j y_i}
\end{array}\right)
\eea

\subsection{Transformations on connections}
Let us define an action of the group of Schlesinger transformations $T_{ij}$ on the space of connections with singularities at $z=z_i.$ As mentioned the action on bundles changes the gluing function. Let us start with the trivial bundle given by the gluing function $1.$ 
The Schlesinger transformation gives another bundle structure, a global section is defined by a pair $S_0,S_{\infty}$ such that $S_{0}=GS_{\infty}$ with $G=G_{ij}G_{\infty}.$ One can define an action on the space of connections as follows: let $\p-A$ be an initial connection in the trivial bundle defined by the same expression over two considered open sets, the transformed connection is  then defined as follows:
\bea
\p-A &&~~\mbox{ over}~~U_{\infty}\nn\\
G(\p-A)G^{-1}&&~~\mbox{over}~~ U_{0}\nn
\eea
After changing basis in $U_{\infty}$ of the form $\widetilde{S_{\infty}}=G_{\infty}S_{\infty}$ we obtain the expression for the connection
\bea
\p-A~\rightarrow~G_{\infty}(\p-A)G_{\infty}^{-1}
\eea
If one changes the trivialization in $U_{0}$ as follows 
$\widetilde{S_{0}}=G_{ij}^{-1}S_{0}$
one obtains
\bea
G(\p-A)G^{-1}~\rightarrow~G_{ij}^{-1}G(\p-A)G^{-1}G_{ij}=G_{\infty}(\p-A)G_{\infty}^{-1}
\eea
Hence the transformed connection can be represented as a global rational connection of the same type as the initial one, the same behavior at $\infty$ is guarantied by the fact that $G_{\infty}$ is holomorphic invertible at $U_{\infty}.$ 

{\Rem 
\label{inv_def} This definition is coherent with the invariant construction, the Hecke transformation on the sheaf language is the operation of taking a subsheaf, defined by a linear condition on values of sections at fixed point. To define an action on the space of singular connections at a singular point one has to choose as a condition the dual covector to one of the eigen direction for the residue of the connection.}

Using the results of the previous section let us calculate the action of the Schlesinger transformation on the connection. 
To observe the normalization condition of $A(z)$ at $\infty$ one should consider the transformation of the form
\bea
\widetilde{G}(z)&=&G_{\infty}^{-1}(\infty) G_{\infty}(z)\nn\\
&=&
\frac 1 {z-1}\left(z-
\left(\begin{array}{cc}
\frac {x_1(y_0-2y_1+x_1 y_0 y_1)}{(1-x_1 y_0)(1-x_1 y_1)} & 
\frac {x_1(1-2 x_1 y_0+x_1 y_1)}{(1-x_1 y_0)(1-x_1 y_1)}\\
\frac {y_0-2y_1+x_1 y_0 y_1}{(1-x_1 y_0)(1-x_1 y_1)} & 
\frac {1-2 x_1 y_0+x_1 y_1}{(1-x_1 y_0)(1-x_1 y_1)}
\end{array}\right)
\right)
\eea

The whole set of Schlesinger transformations for the $3$-point case relevant to the analysis of the Painlev\'e VI equation was calculated in \cite{MS}

\subsection{Action on Bethe vectors}
Let us introduce the notation $Z=\{z_i\},~W=\{w_j\},~\Lambda=\{\l_i\}.$ Let us also introduce symbols for the following objects:
\begin{itemize}
	\item $OS(Z,W,\Lambda)$ - a solution for the system of Bethe equations, that is  the set $\m_i$ which solves the system \ref{b} for the fixed points (the poles of the Lax operator) $z_1,\ldots,z_k$ and  
	$w_1,\ldots,w_{k-2}$ and the corresponding highest weights $\l_1,\ldots,\l_k$ and $1,\ldots,1$; 
	\item $OL(Z,W,\Lambda)$ - a solution of the linear problem \ref{baxter} of the form \ref{SL} with the given set of parameters;
	\item $ML(Z,W,\Lambda)$ - a solution of the matrix linear problem \ref{aux} of the form \ref{ML} with the given set of parameters, it has two components $ML(Z,W,\Lambda)_{1,2}$;
\end{itemize}
{\Rem The choice of the highest weights $1$ at ``moving poles'' $w_i$ is not necessary but in a sense generic. One can take the potential 
\bea
U=\sum_{j=1}^{m}\frac {-1/4(\eta_j+2)\eta_j}{(z-w_j)^2}+\sum_{i=1}^{k}\frac {1/4+det A_i}{(z-z_i)^2}+
\sum_{j=1}^{k-2}\frac {H_{w_j}}{z-w_j}+\sum_{i=1}^{k}\frac {H_{z_i}}{z-z_i}
\eea
with generic values of the highest weights . This realizes if one demand $a_{12}(z)$ to have  zeros $w_j$ with multiplicities $\eta_j$ subject to the condition $\sum_{j=1}^m \eta_j=k-2.$}

~\\
The results of our previous considerations can be summarized in the following: for a solution of the Bethe system $OS(Z,W,\Lambda)$ one can consider a solution $ML(Z,W,\Lambda)$ in virtue of the theorem \ref{back}. Then one can define a family of transformations:
\begin{itemize}
	\item {\bf Dual solution~} Symbolically this transformation expresses vie the diagram
	\bea
	\begin{array}{ccccccccc}
	OS(Z,W,\Lambda) & &  & & & & OS(Z,\widetilde{W},\Lambda)\\
	\downarrow & & & & & & \uparrow \\
	OL(Z,W,\Lambda) & \rightarrow & ML_1(Z,W,\Lambda) & \rightarrow 
	& ML_2(Z,W,\Lambda) & \rightarrow &  OL(Z,\widetilde{W},\Lambda)	
	\end{array}\nn
	\eea
	\item {\bf Schlesinger transformation~} Using the notation below one has 
	\bea
	\begin{array}{ccccccccc}
	OS(Z,W,\Lambda) & &  & & && OS(Z,\widetilde{W},\widetilde{\Lambda})\\
	\downarrow & & & &  & & \uparrow \\
	OL(Z,W,\Lambda) & \rightarrow & ML(Z,W,\Lambda) & \rightarrow 
	& ML(Z,\widetilde{W},\widetilde{\Lambda}) & \rightarrow &  OL(Z,\widetilde{W},\widetilde{\Lambda})	
	\end{array}\nn
	\eea
Depending on the choice of the subspace of the upper and lower Schlesinger transformations one obtains the following action of $T_{ij}$ on the set of highest weights
	\bea
	(\ldots,\l_i,\ldots,\l_j,\ldots)&\longmapsto &(\ldots,\l_i+1,\ldots,\l_j-1,\ldots)\nn\\
	(\ldots,\l_i,\ldots,\l_j,\ldots)&\longmapsto &(\ldots,\l_i+1,\ldots,\l_j+1,\ldots)\nn\\
	(\ldots,\l_i,\ldots,\l_j,\ldots)&\longmapsto &(\ldots,\l_i-1,\ldots,\l_j-1,\ldots)\nn\\
	(\ldots,\l_i,\ldots,\l_j,\ldots)&\longmapsto &(\ldots,\l_i-1,\ldots,\l_j+1,\ldots)\nn
	\nn\eea 
\end{itemize}

\section*{Discussion}
\subsection*{Painlev\'e equations}
The idea of discreet transformations preserving the condition of ``minimality'' of the monodromy exploited here lies in the middle of the theory of isomonodromic transformations. The same group of transformations act on the space of solutions of the Painlev\'e VI equation realized as the monodromy preserving condition for the Fuchsian system (the references can be found for example in \cite{GP}):
\bea
(\p-A(z))\Psi(z)=0\nn
\eea
where 
\bea
A(z)=\left(
\begin{array}{cc}
a_{11}(z) & a_{12}(z) \\
a_{21}(z) & a_{22}(z)
\end{array}\right)=\frac {A_0}{z}+\frac {A_1}{z-1}+\frac {A_t}{z-t} \nn
\eea
One can expect that the problem of continuous deformation of the spectrum for the quantum Gaudin model with respect to the parameters of the model - the marked points, is equivalent to the Schlesinger system of equations, hence for some particular case of the model is just the Painlev\'e VI equation. This is not a priori evident because of the essentially resonant case of the connection appearing in the isomonodromic formulation of the Bethe equation for the Gaudin model.

\subsection*{Langlands correspondence}
The quantum Gaudin model plays an important role in the geometric Langlands correspondence \cite{Fr,CT06-1}, namely the correspondence is provided by the relation: 
\vskip 2mm
\centerline{``Gaudin-Hitchin'' $D$-module $\Leftrightarrow$ flat connection ($G$-oper)}
\smallskip
In the arithmetic case moreover one has an important property of coincidence of the corresponding $L$-functions. In the geometric case over $\C$ there is no natural notion of the Frobenius automorphism and hence of the Galois-side $L$-function. Presumably, the symmetry elaborated here is related namely to the Galois side of the correspondence.

Recently in \cite{GLO} it was observed another appearence of the arithmetic $L$-functions in quantum integrable systems, namely in the theory of quantum affine Toda chain.

\end{document}